# Power Synthesis of Maximally-Sparse Linear Arrays Radiating Shaped Patterns through a Compressive-Sensing Driven Strategy

Andrea Francesco Morabito, Antonia Rita Laganà, Gino Sorbello, and Tommaso Isernia

*Abstract*—We present an innovative approach to the synthesis of linear arrays having the least possible number of elements while radiating shaped beams lying in completely *arbitrary* power masks. The approach, based on theory and procedures lend from Compressive Sensing, has two innovative key features. First, it exploits at best the multiplicity of equivalent *field* solutions corresponding to the many different *power* patterns lying in the given mask. Second, it a-priori optimizes those parameters that affect the performance of Compressive Sensing. The overall problem is formulated as two convex programming routines plus one local optimization, with the inherent advantages in terms of computational time and solutions' optimality. An extensive numerical comparison against state-of-the-art procedures proves the effectiveness of the approach.

## I. INTRODUCTION

The synthesis of non-uniformly spaced arrays [1] is of interest in a large number of applications. While the first design techniques date back to the Sixties (see for instance [1],[2]), the problem has been the subject of a renewed interest in the last years [3]-[20]. In particular, attention has been devoted to the synthesis of 'maximally-sparse' arrays [6]-[15], i.e., arrays fulfilling assigned mission requirements by exploiting the minimum number of active elements. Many different solution approaches have been proposed to deal with such problem, recently including global optimization [15], the Forward Backward Matrix Pencil Method (FBMPM) [11]-[13], and the Linear Programming (LP) procedure proposed in [14]. Finally, an important body of work is given by a number of techniques based on the Compressive Sensing (CS) theory [6]-[10].

On the other side, a brief look to the characteristics of the different approaches shows that there is indeed room for improvement, as detailed below.

First, it is interesting to note that the MPM and FBMPM, although being very effective, have some theoretical flaw (as the authors themselves state at the end the introduction of [12]). In fact, in order to find the locations of the array elements, the MPM requires that the poles of the equation identified by organizing the desired pattern data as an Hankel matrix and performing its singular value decomposition lie on the unit circle in the complex plane [11]. However, this property usually does not hold true in the case of shaped beams, this being the main reason of replacing the MPM with the FBMPM in the synthesis of such kind of fields [12]. On the other side, as stated by authors, the additional relationships enforced by FBMPM, although being very useful, represent *a necessary* but not *a sufficient* condition in order to guarantee the unitary amplitude of the mentioned poles [12], so that it is still possible to further improve the achieved results.

Second, both the FBMPM and the existing CS-based approaches aim to fit at best *both the amplitude and phase* distributions of a reference far-field pattern. Such a circumstance completely ignores the fact in array antennas synthesis one is generally interested in *power* distributions, rather than *amplitude and phase* far-field distributions. Obviously, *a priori* establishing given phase distributions implies a reduction of the degrees of freedom available to the designer, thus reducing the number of possible solutions [21]. The approach in [14] apparently overcomes such a limitation, as a formulation is given in terms of a 'mask' for the pattern. Unfortunately, this technique is based on the assumption that the radiated field is a *real* function, while it is known that in case of shaped beams the optimal far-field pattern, i.e., the one with the least dimension for given performances, is in general a complex function [21]. Hence (see Sections II and IV), a number of possibly convenient solutions are hidden to the designer also in such a case.

In the following, we propose, discuss and test a new approach to the synthesis of shaped patterns by means of (maximally) sparse linear arrays. The approach is able to take into account the multiplicity of possible fields able to fulfill given upper and lower bounds on the power pattern. Although developed for the case of linear arrays, the approach can be extended to the case of patterns requiring a circular symmetry, and it can be used in the case of two-dimensional factorable patterns, which are of actual interest in several applications.

The paper is organized as follows. In Section II we recall basic results of CS and discuss its strengths and weaknesses in relationship with synthesis problems. Moreover, we also briefly recall useful results already available in the synthesis of shaped beams, with particular emphasis on the case of linear arrays. The proposed approach is presented and assessed in Sections III and IV, respectively, the latter providing successful numerical comparisons with all methods and results in [7],[10],[12],[14]. Conclusions follow.

## II. A COUPLE OF BASIC RESULTS

In order to better clarify the usefulness of CS in synthesis problems, as well as the meaning and the interest of the approach we are proposing, in this Section we briefly review a couple of general theoretical results respectively related to the CS theory and to the synthesis of shaped beams.

*II.1 Interest and weaknesses of Compressive Sensing in synthesis problems*

The well-known circumstance that the aperture source and its far-field distribution are essentially related by a Fourier Transform constitutes a formidable basis in order to apply to array synthesis all theories and procedures arising from [22] and related (see for instance [23],[24]).

Roughly speaking, the measurements one would use in the recovery of a sparse signal are substituted by the far-field specifications, while the array layout and excitation represent the sparse signal to be 'recovered' through CS [6]-[10].

In applying CS to such a problem, the common approach is that of looking for a fitting of *both the amplitude and phase* distributions of a reference far-field pattern, and the problem's actual unknowns are both the array elements excitations and locations. In pursuing such a goal, inspired from [22]-[24], the crucial antenna-design operation is usually that of minimizing the (weighted) $\ell_1$ norm of excitations [6]-[10], as defined on a very dense auxiliary grid. One may argue that both terms 'recovery' and 'Compressive Sensing' are questionable, as one cannot be sure that a sparse array able to realize a given pattern certainly exists, so that one could talk in terms of 'sparsity enhancing' rather than of 'Compressive Sensing'. On the other side, if a sparse array able to fulfill the assigned technical requirements exists then the CS theory provides the rigorous conditions and the actual algorithms allowing to unveil it, so that the adopted terminology (and framework) makes indeed sense.

The CS-based procedures in [6]-[10] show excellent capabilities for the synthesis of sparse arrays. On the other hand, they also have a number of drawbacks, and it is indeed the CS theory which allows understanding weaknesses and developing a number of improvements.

First, while the CS theory provides theoretical conditions and practical procedures for a faithful recovery in case of linear problems, it must be stressed that a power-pattern synthesis problem, even when looking for a generic (non-sparse) source distribution, is indeed a *non-linear* problem. In fact, as already stated, one is generally interested in *power* distributions, rather than *amplitude and phase* far-field distributions, and a-priori establishing given phase distributions, as it is done in the present CS-based procedures, implies a reduction of the degrees of freedom available to the designer, thus hiding possible solutions [21].

As a consequence, one should either develop a CS theory for non-linear problems, or devise possible ways to take advantage from the fact that many different complex patterns could equally fulfill the given power pattern specifications (which is our choice, see Section III).

A second weakness of present procedures can be understood from a number of results including *Theorem 1.3* in [22] and/or condition (6) in [23]. In these references, it is shown that the recovery of a signal having an original dimension equal to $N$ and a 'sparse' or 'compressed' size equal to $S$ results successful with overwhelming probability only if the number of independent measurement (say $M$) exceeds a threshold which grows with both $S$ and $\log N$ [22]-[24].

Now, in the CS application to array synthesis, $M$, $N$, and $S$ respectively map to the cardinalities of *far-field* specifications, *initial* array layout (including elements with zero excitation), and *final* sparse array layout (counting only active elements) [6]-[10]. As a consequence, since any field radiated by a finite-dimensional source has a finite number of *degrees of freedom* (and hence a limited value of $M$) [25], a very dense original array layout (and hence a very large value of $N$) may imply a violation of the conditions under which CS guarantees an effective recovery, i.e., an effective array synthesis.

As a consequence, one should be careful about the choice of $N$, i.e., of the initial grid wherein CS is eventually applied, while still allowing in the overall procedure to locate the active antennas anywhere.

*II.2 A theoretical result in the synthesis of linear sources*

A number of procedures for the synthesis of shaped beams through standard 'non-sparse' arrays (such as the Woodward Lawson technique [26] and Gradient Search and Conjugate Matching) are reviewed in [20]. Another relevant and widely diffused approach to synthesis (which includes conjugate matching) is the so-called Alternating Projection technique [27].

On the other side, all these synthesis methods do not take into account that the synthesis of sources able to realize *power* patterns lying within a given mask is indeed a *non-linear* and *non-convex* problem. As a consequence, the final result of all the above synthesis methods depends indeed on the initial assumptions or on the starting point of the adopted algorithm. In fact, an exceptionally large number of procedures based on global optimization procedures has been devised (see [28],[29] and references therein).

In case of equispaced linear arrays a completely different approach, which overcomes the non-convexity of the problem as well as the computational burden (and weaknesses [30]) associated to global optimization, is possible [21]. The approach can be naturally extended to the case of continuous sources by considering the limiting case of vanishing small spacings (and adding constraints on the invisible part of the spectrum) [31].

Such an alternative approach is based on the circumstance that the square-amplitude array factor of a linear equispaced array of $Q$ antennas can be expressed as

$$P(u) = \sum_{q=-Q+1}^{Q-1} D_q e^{jqu} \qquad (1)$$

with $u=\beta d\sin\alpha$ and $\beta=2\pi/\lambda$, $\lambda$, $\alpha$, and $d$ being the wavelength, the angle between the boresight and observation directions, and the spacing between adjacent elements, respectively. Since $P$ must be a real and non-negative function, the conditions

$$D_q = D_{-q}^* \qquad q = 1, 2, ..., Q-1 \qquad (2)$$

where * means complex conjugation, and

$$\sum_{q=-Q+1}^{Q-1} D_q e^{jqu} \geq 0 \qquad (3)$$

must hold true.

Hence, one can identify a power pattern fulfilling given shape constraints by solving the following optimization problem:

$$\begin{cases} \sum_{q=-Q+1}^{Q-1} D_q e^{jqu} \leq UB(u) & (4.a) \\ \sum_{q=-Q+1}^{Q-1} D_q e^{jqu} \geq LB(u) & (4.b) \\ D_q = D_{-q}^* \quad q=1,2,...,Q-1 & (4.c) \\ \sum_{q=-Q+1}^{Q-1} D_q e^{jqu} \geq 0 & (4.d) \end{cases}$$

wherein the arbitrary (real and non-negative) functions $UB$ and $LB$ respectively denote the upper and lower bounds exploited to shape the field amplitude according to the particular application at hand. By taking into account the band-limitedness of $P$, expressions (4) can be substituted with a sufficiently fine discretization [21],[25] so that it can be seen as a system of ordinary linear inequalities in the $D_q$ coefficients. Notably, such a problem can be easily solved by using any LP tool.

Then, representation (1) can be factored in two polynomials being the complex conjugate each of the other, which can be interpreted as the array factor and its complex conjugate. As a consequence, one is able to solve the shaped beam synthesis problem through solution of (4) and suitable factorization techniques [21]. Notably, more than a single solution can be found, as the factorization of the reference power pattern $P$ in complex conjugate factors is generally non-unique, so that the same *power* pattern corresponds to a (generally large) number of different *far-field* distributions. All of the solution can be easily generated, at least in principle, by means of the so called 'zero-flipping' operation [21].

With respect to the specific subject of this paper, the above result has a two-fold usefulness. First, it offers a natural way to translate *power-pattern* specifications into *field* specifications, thus allowing to exploit the standard (linear) CS framework. Second, the consideration of all possible field determinations allows to keep under control (and take advantage from) the multiplicity of solutions available for each fixed power pattern.

### III. THE PROPOSED APPROACH TO SYNTHESIS

The aim of the procedure is to perform the *power* synthesis of a shaped beam lying in a *fixed and arbitrary* mask through an array antenna composed by the *minimum number of elements*. By taking into account limitations discussed in Section II.1, and taking advantage from the result recalled in Section II.2, the power synthesis problem at hand has been conveniently formulated as the following three-steps procedure.

In the **first step**, determine some convenient nominal power pattern to be pursued, and find all the different field patterns corresponding to it.

In the **second step**, by taking advantage from the field patterns determined in step one, use CS procedures in order to synthesize a sparse array having the minimum possible number of elements.

The above two steps allow taking relevant advantage from the fact that a single power pattern corresponds to a large number of different patterns, but they still do not exploit all the available degrees of freedom. In fact, there may be many different power patterns lying in the same mask.

The **third step** is hence devoted to perform a refinement of the solution by taking advantage from this circumstance. In particular, modifications of the initial power pattern are pursued in such a way to realize a further sparsification of the array layout. Obviously, the original mask is still enforced on the on the final power pattern.

Details on the three steps are respectively given in the three following subsections.

*III.1. Step 1: Exploiting the multiplicity of solutions for a given power pattern*

Let us suppose a solution to (4) has been found. Then, by introducing the auxiliary variable $z=e^{ju}$, and exploiting the Fundamental Theorem of Algebra, representation (1) can be factored in terms of its zeroes. Their proper clustering allows the factorization of the reference power pattern $P$ in complex conjugate factors. Notably, one will have many different clustering possibilities corresponding to number of different far-field distributions. To express this multiplicity of solutions, let us consider the matrix

$$\underline{\underline{A}} = \begin{bmatrix} \underline{a}_1 \\ \underline{a}_2 \\ \vdots \\ \underline{a}_W \end{bmatrix} \qquad (5)$$

where $W$ is the overall number of different array factors corresponding to $P$ and the vector $\underline{a}_w=[a_{w,1},a_{w,2},...,a_{w,Q}]$ denotes the excitation vector corresponding to the w-th array factor for $w=1,2,...,W$.

As discussed in [21], the number of equivalent solutions is $W=2^{R'/2}$, $R'$ being the number of zeroes of $P$ which do not lie on the unit circle. As *W* grows very rapidly with *R'*, it makes sense to discuss the choice of the parameters *Q* and *d*, which determine the total number of roots of *P*, i.e., *R*=2*Q*-2, and the reference-array size, i.e., *T*=(*Q*-1)*d*.

Due to the bandlimitedness of the fields radiated by non-superdirective sources [25],[31], the actual feasibility or unfeasibility of a power pattern mask is essentially determined by the antenna's electrical size *T*/λ. In fact, for a fixed *T*, increasing *Q* means increasing *R*, thus apparently furnishing a larger number of degrees of freedom. On the other side, increasing *Q* also means to reduce the array elements spacing. The reduction of *d*, in turn, forces the user to add constraints in the invisible part of the spectrum in order to avoid superdirective sources [21],[31]. As a consequence, the additional roots which are apparently available must be used to bind the antenna's reactive energy (and are indeed expected to lie on the unit circle). Therefore, provided *d*≤λ/2, different values of *Q* and *d* corresponding to the same value of *T* are expected to lead to very similar radiation performances in the visible part of the spectrum. We have verified this circumstance in a large set of synthesis problems, solving each of them by choosing *d*=0.5λ, *d*=0.4λ, and *d*=0.3λ. At the end of the overall proposed approach, they all led to identical results in terms of both radiation performance and sparse array's final number of elements.

Let us discuss now the choice, amongst all the equivalent solutions (5), of the field to be pursued through CS. A first possibility could be that of applying a CS-based procedure to all the different fields arising from the factorization step, and then to observe the resulting number of active array elements. A similar strategy has been adopted in [12] by Liu and co-workers which, starting from a power solution in [21], have applied the FBMPM to all the equivalent *field* solutions.

A second possibility could be that of *a-priori* selecting (by exploiting some convenient ad-hoc criterion) one amongst the different field distributions fulfilling the initial constraints. This strategy, which is the one exploited by our approach, may allow avoiding a heavy computational burden in those test cases having a very large number of equivalent solutions.

Following the CS-theory [22]-[24], an intuitive choice is to consider the excitations set corresponding to the array layout with the *maximum amount of sparsity*. Therefore, amongst all the available far-field distributions, a convenient choice is to select

$$f_r(\alpha) = \sum_{q=1}^{Q} a_{r,q} e^{jq\beta d \sin\alpha} \quad (6)$$

where $f_r(u)f_r^*(u)=P(u)$, and

$$\|\underline{a}_r\|_1 \leq \|\underline{a}_w\|_1 \quad w=1,2,...,W \quad (7)$$

as *reference* field for step 2 of our procedure.
In fact, according to the CS-theory concerning relaxation of $\ell_0$-norm optimization problems, $f_r$ represents, by construction, the far-field distribution that more easily lends itself to a 'sparsification' process (see [22]-[24] for further details).

As discussed with the help of some examples in the Section IV, application of (7) works very well even in those cases where *d* is chosen in the order of λ/2, which can be attributed again to the finite number of *degrees of freedom* of finite-dimensional sources [25].

*III.2. Step 2: CS-driven engine to recover a preliminary array layout*

Once the 'optimal' reference *far-field* distribution has been identified, we apply a CS-inspired procedure to synthesize it. We denote with $f_b$ the optimized far field and with the vectors *x* and *b* the locations and excitations of the *N*-elements linear array radiating it, respectively.

Many different procedures are commonly used in sparse recovery, such as greedy approaches [32], iterative thresholding [33], Basis Pursuit [34] or Least Absolute Shrinkage and Selection Operator (LASSO) [35]. Taking into account the theoretical results presented in [22], we adopt an approach based on a $\ell_1$ relaxation of the original problem, which is similar in spirit to the LASSO approach. In particular, we formulate the problem as follows:

$$\min_{\underline{b}} \|\underline{b}\|_1 \quad (8.a)$$

subject to

$$\begin{cases} |f_b(\alpha)|^2 \leq g(\alpha) \quad \forall \alpha \in \tau_1 & (8.b) \\ \dfrac{\|f_b(\alpha) - f_r(\alpha)\|_2}{\|f_r(\alpha)\|_2} \leq \varepsilon \quad \forall \alpha \in \tau_2 & (8.c) \end{cases}$$

wherein

$$f_b(\alpha) = \sum_{n=1}^{N} b_n e^{j\beta x_n \sin\alpha} \quad (9)$$

Notably, the problem is reduced in such a way to a Convex Programming (CP) optimization. In such a formulation:

- the cost function (8.a) is based on CS theory and has the same aim of the functional minimized in [7]-[10], i.e., to induce a minimization of the arrays' active elements number;

- constraint (8.b) is devoted to control the sidelobes amplitude and enforces an upper bound $g(\alpha)$ on the synthesized power pattern in the angular region denoted with $\tau_1$. Note *g* can be an arbitrary (real and positive) function so that field 'notches' can be eventually realized;

- constraint (8.c) ensures that, in the angular region denoted by $\tau_2$, the optimized array factor fits the reference one with a precision ε.

Notably, given its mathematical formulation, provided an intersection between constraints (8.b) and (8.c) does exist, problem (8) provides always a unique solution provably representing the global optimum. Moreover, provided conditions for the validity of CS sparse recovery hold true, such a solution also will be a *maximally sparse* solution to the synthesis problem [6]-[10].

As a final comment on this step, let us discuss the choice of $N$ and $\underline{x}$ in (9). Once the array aperture size has been fixed on the basis of the particular radiation and geometrical specifications, following the theory in [6]-[10], we define $\underline{x}$ as a vector uniformly spanning it. Note that, in order to optimize the CS performance, we use values of $N$ considerably smaller than those adopted in previous procedures. Such a choice (see also Section II.1) will boost potentialities of CS from both point of views of speed of convergence and optimality of final results [22]-[24]. On the other side, possible accuracy losses in identifying the array locations, deriving from choosing in such a step an elements spacing larger than in other approaches, are then recovered in the third step of the design procedure.

*III.3. Step 3: Further sparsification and solution refinement*

The last part of the synthesis is devoted to a further 'sparsification' and optimization of the array layout.

In the first part of it, which is similar in spirit to the so-called iterative thresholding approaches [33], we further reduce the number of array elements by:

1) discarding the antennas having an excitation amplitude lower than a threshold $\upsilon$;

2) recursively substituting each couple of remaining elements whose distance is lower than a threshold $\sigma$ with a single element placed in the middle point between them.

In the following, the elements number and the locations set coming out from these operations are denoted with $S$ and $\underline{y}$, respectively, while a sketch illustrating such modifications is given in Fig. 1.

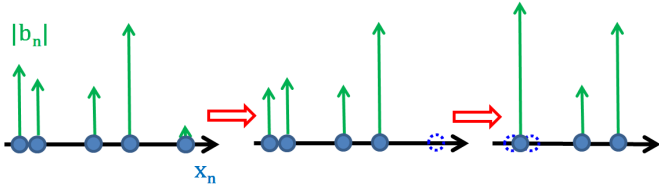

Fig. 1. Third step of the design procedure: sketch concerning the sparsification of the array layout coming out from step 3.

Then, the solution is further improved by determining slight shifts on array locations (and the optimal excitations associated to them) so as to improve the overall radiation performances without increasing the number of elements.

In particular, we look for vectors $\underline{c}$ and $\underline{\Delta y}$ such that the final far field distribution

$$f_{y1}(\alpha) = \sum_{s=1}^{S} c_s e^{j\beta(y_s + \Delta y_s)\sin\alpha} \quad (10)$$

is given by the solution the following optimization problem:

$$\min_{\underline{c}, \underline{\Delta y}} \frac{\|f_{y1}(\alpha) - f_r(\alpha)\|_2}{\|f_r(\alpha)\|_2} \quad \forall \alpha \in \tau_2 \quad (11.a)$$

subject to

$$|f_{y1}(\alpha)|^2 \leq g(\alpha) \quad \forall \alpha \in \tau_1 \quad (11.b)$$

In fact, provided that the minimum value of the functional in (11.a) does not result larger than $\varepsilon$, solving problem (11) guarantees a radiation performance equivalent to the one coming out from problem (8) while saving $N-S$ array elements.

Unfortunately, due to the way the unknown $\underline{\Delta y}$ appears in (10), the optimization problem (11) is a non-convex one. On the other side, one is just looking for 'small' shifts, so that (10) can be reasonably approximated by

$$f_{y2}(\alpha) = \sum_{s=1}^{S} c_s e^{j\beta y_s \sin\alpha}(1 + j\beta\Delta y_s \sin\alpha) \quad (12)$$

As a consequence, the last step can be formulated as in (11) by using $f_{y2}$ instead of $f_{y1}$ and adding the convex constraint

$$-\eta \leq \beta\Delta y_s \leq \eta \quad \forall s \quad (13)$$

being $\eta$ a user-defined constant such that $0<\eta<<1$.

Notably, even if the optimization problem, unlike (4) and (8), is still not of the CP class, it is nearly so by virtue of (13), so that occurrence of sub-optimal solutions can be avoided while local optimization procedures can still be used. Finally, the adopted formulation allows the addition of the convex constraint

$$[(y_{S+1} + \Delta y_{S+1}) - (y_S + \Delta y_S)] \geq \psi \quad s = 1,2,\ldots,S-1 \quad (14)$$

as a powerful way to avoid a too small spacing between adjacent elements (possibly leading to undesired mutual-coupling effects).

It is worth noting that this last step allows counteracting possible drawbacks deriving from a value of $N$ smaller than in other CS-based approaches, and compensating for possible mismatches deriving from the 'discarding' and 'unification' of elements described above. Moreover, it pursues a field fitting just in the main-beam zone, while upper bounds are used for the sidelobes regions. As already stated, the possibility of modifying the power pattern with respect to the nominal one fixed in step 1 allows recovering a significant number of degrees of freedom with respect to more usual approaches.

IV. NUMERICAL ASSESSMENT

In order to assess the actual capabilities of the proposed approach, we have compared its outcomes with those of the state-of-the-art techniques available for the synthesis of shaped beams through 'maximally-sparse' arrays.

In order to provide a comparison as complete as possible, we discuss in the following six test cases. In particular, the first two examples respectively concern a comparison with two techniques which tackle the problem at hand without exploiting the CS theory, namely the LP procedure presented in [14] and the FBMPM introduced in [12]. Then, in the third, fourth, and fifth numerical examples two state-of-the-art techniques exploiting the CS theory, namely the Bayesian Compressive Sensing (BCS)-based procedure described in [7] and CS-based weighted-norm approach proposed in [10], are considered. Finally, the sixth test case concerns a simple extension of the approach to the synthesis of a planar array. By the sake of clarity, we resume in Table 1 the main parameters used for the different linear arrays.

| Test case | Cfr. vs. [14] (flat-top) | Cfr. vs. [12] (cosecant) | Cfr. vs. [7] (flat-top) | Cfr. vs. [7] (cosecant) | Cfr. vs. [10] (cosecant) |
|---|---|---|---|---|---|
| $W$ (Number of equivalent solutions) | 512 | 32 | 16 | 32 | 32 |
| $\varepsilon$ (Fitting precision inside $\tau_2$ region) | $2\times10^{-2}$ | $17\times10^{-2}$ | $5\times10^{-3}$ | $17\times10^{-2}$ | $22\times10^{-2}$ |
| $\tau_2$ (Fitting region) | $|\sin\alpha|\leq 0.37$ | $-0.05\leq\sin\alpha\leq 0.85$ | $|\sin\alpha|\leq 0.31$ | $0\leq\sin\alpha\leq 0.83$ | $-0.18\leq\sin\alpha\leq 0.59$ |
| $L/\lambda$ (Size) | 11.68 | 7.28 | 6.80 | 6.93 | 7.53 |
| $\psi/\lambda$ (Minimum elements spacing) | 0.40 | 0.48 | 0.46 | 0.51 | 0.55 |
| $S$ (Final elements number) | 13 | 11 | 10 | 12 | 12 |
| Elements number saving vs. best case available in literature | 58% | 15% | 9% | 8% | 8% |

Tab. I. Simulation parameters adopted for the test cases concerning one-dimensional arrays.

Notably, by virtue of the extremely low computational burden of the overall design algorithm, each example required less than 15 seconds to be generated by a calculator having an Intel Core i7-3537U 2.50GHz CPU and a 10 GB RAM.

Finally, it is worth noting that, although the proposed procedure can manage any element factor in the synthesis, all the following power-pattern figures just refer to the square-amplitude distribution of the array factors.

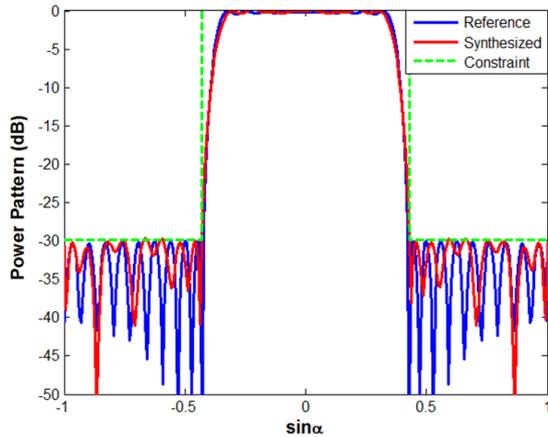

Fig. 2. Reference (blue curve, from [14]) and synthesized (red curve) power pattern in the test case of Susection IV.1. Exploited upper-bound constraint (green curve) is also depicted. Simulation parameters: $d/\lambda=0.5$; $Q=29$; $N=201$; $v=0.05$; $\sigma/\lambda=0.1$; $\eta/\beta=0.025$.

*IV.1 Comparison with [14] (flat-top beam)*

We adopted as reference the power pattern depicted in blue color in Fig. 2, which has been synthesized in [14] by means of a 31-elements sparse array. This flat-top field guarantees a ripple equal to 0.4455 dB for $-20°\leq\alpha\leq 20°$, and fulfills an upper-bound constraint (depicted in green color in Fig. 2) of -30 dB for $-90°\leq\alpha\leq -25°$ and $25°\leq\alpha\leq 90°$. By factorizing this power pattern, we have been able to determine 512 equivalent field solutions. Then, we have selected amongst them the one corresponding to the excitation distribution having the minimum $\ell_1$ norm and, by applying to the latter the proposed approach, we identified a 13-elements sparse array able to radiate the power-pattern distribution depicted in red color in Fig. 2. As it can be seen, the synthesized solution results equivalent, in terms of radiation performance, to the reference one. Therefore, the proposed procedure has allowed, with respect to the best solution shown in [14], a 58% elements saving without experiencing any power-pattern worsening. The synthesized sparse array's excitation amplitudes and phases are respectively depicted in figures 3 and 4.

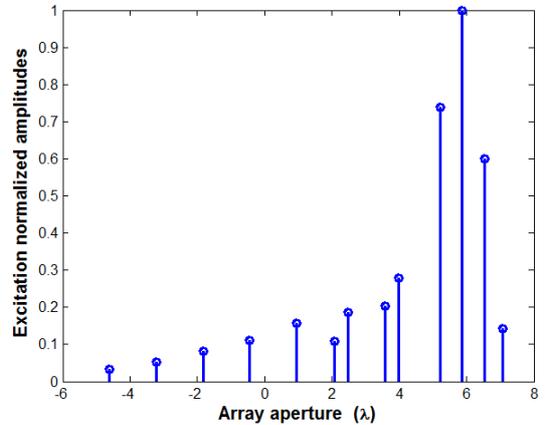

Fig. 3. Excitation amplitudes of the maximally-sparse array radiating the red pattern of Fig. 2.

*IV.2 Comparison with [12] (cosecant beam)*

We used as reference square-amplitude array factor the power pattern depicted in blue color in Fig. 5.
In [12], the FBMPM allowed generating it thorugh a sparse array composed by 13 elements located over an aperture of $7.5\lambda$. By exploiting the proposed approach, we have been able to generate the power pattern depicted in red color in Fig. 9 by exploiting a sparse array composed by 11 elements located over an aperture of $7.4\lambda$. Therefore, without experiencing any loss in terms of radiation performance, and exploiting a practically equivalent aperture size, we reduced of the 15.4% the elements number achieved in [12]. The amplitude and phase distributions of the synthesized array excitations are shown in figures 6 and 7, respectively.

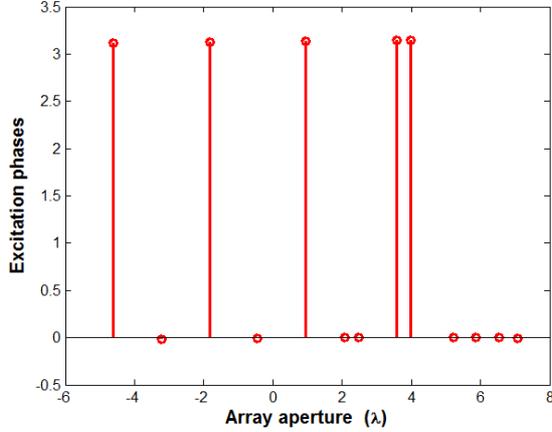

Fig. 4. Excitation phases of the maximally-sparse array radiating the red pattern of Fig. 2.

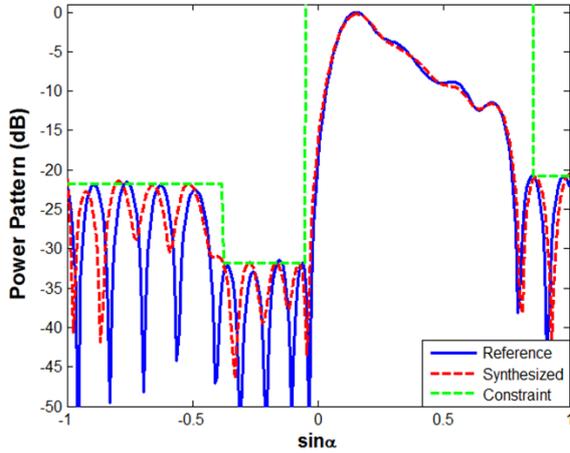

Fig. 5. Adopted upper bound (green curve) and reference (blue curve, from [12]) and synthesized (red curve) power patterns for the second test case. Simulation parameters: $d/\lambda=0.5$; $Q=16$; $N=321$; $\nu=0.005$; $\sigma/\lambda=0.025$; $\eta/\beta=0.05$.

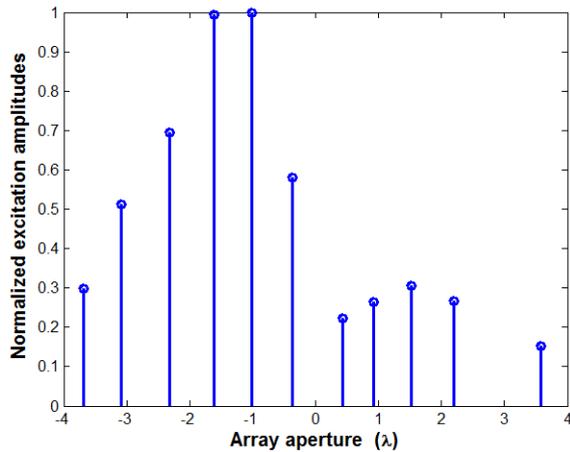

Fig. 6. Excitation amplitudes of the maximally-sparse array radiating the red pattern of Fig. 5.

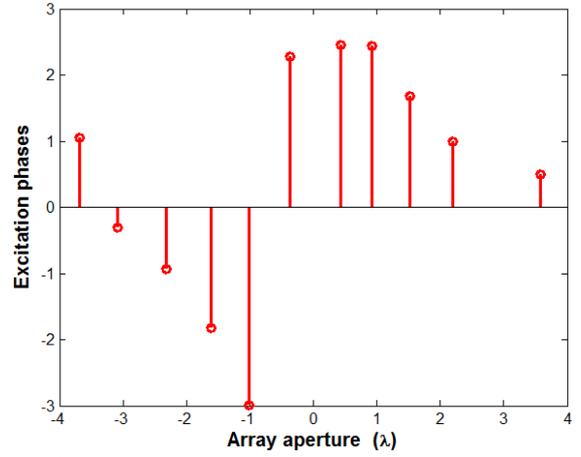

Fig. 7. Excitation phases of the maximally-sparse array radiating the red pattern of Fig. 5.

*IV.3 Comparison with [7] (flat-top beam)*

In the third test case, we used as reference power pattern the flat-top beam depicted in blue color in Fig. 8, which has a maximum ripple of ±0.58 dB, a peak sidelobe level of -35.6 dB with respect to the maximum power pattern value for $|\alpha|\geq 27.5°$. In [7], BCS has been effectively exploited to generate it through a sparse array composed by 11 elements located over an aperture of $7\lambda$. Also, this power pattern has been generated in [12] by exploiting the FBMPM and using 10 radiating elements.

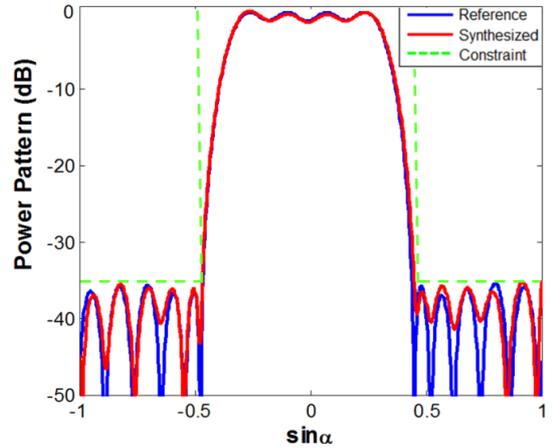

Fig. 8. Third test case: adopted upper bound (green curve); reference (blue curve, from [7]) and synthesized (red curve) power patterns. Simulation parameters: $d/\lambda=0.5$; $Q=15$; $N=38$; $\nu=0.04$; $\sigma/\lambda=0.19$; $\eta/\beta=0.09$.

As first design operation, we have generated this power pattern without changing any parameter with respect to [7] (see Table 1). Then, we have factorized the resulting polynomial and applied our CS-based routine to all the equivalent far-field distributions corresponding to it, with the aim of assessing the effectiveness of the adoption of (7). In particular, in each instance we have analyzed the $\ell_1$ norm of the excitation set of the reference array factor and counted the active elements of the sparse array coming out from the CS routine. The outcomes of such experiments have been summarized in Fig. 9. As it can be seen, the excitation set having the minimum $\ell_1$ norm is also the one that leads to the minimum number of active elements in the array designed through CS. This circumstance (which has been also verified in the previous and following test cases) confirms the arguments in Section III.1 about the capability of identifying in a simple fashion the far-field distribution which more easily lends itself to a CS-based 'sparsification' process.

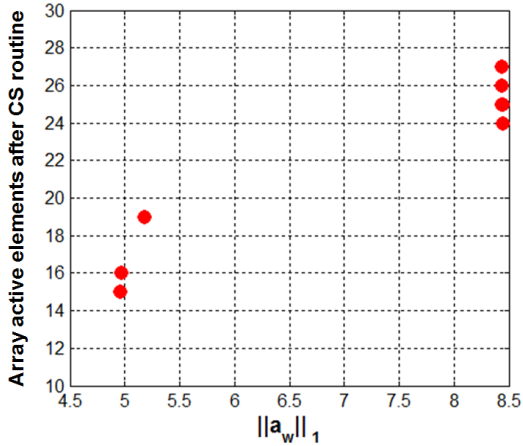

Fig. 9. Concerning the impact on the CS performance of the $\ell_1$ norm of the excitations of the different solutions coming out from the statement of the reference power pattern.

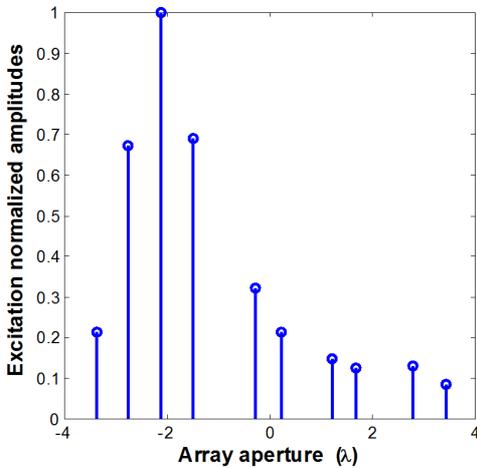

Fig. 10. Excitation amplitudes corresponding to the red pattern of Fig. 8.

By applying the post-processing procedure to the array layout having the minimum number of elements at the end of the CS routine, we have achieved a radiation performance which favorably compares to [7] through a sparse array composed by 10 elements located over an aperture of 6.8λ with a minimum spacing of 0.46λ.

The achieved array locations and excitations are depicted in Fig. 10 (amplitude distribution) and Fig. 11 (phase distribution), respectively, while Fig. 8 shows the exploited upper-bound function and a superposition of the reference and synthesized square-amplitude array factors. As it can be seen, the achieved power pattern perfectly fits the reference one.

Notably, in this test case the results achieved by means of the proposed strategy are equivalent to those one reported in [12], while they turn out being improved with respect to [7] in all terms of number of elements, dimensions and performances. Finally, we highlight the important circumstance that the execution of the third step of the procedure allowed us to reduce $N$ of about the 50% with respect to the case wherein just the first two steps of the approach are exploited.

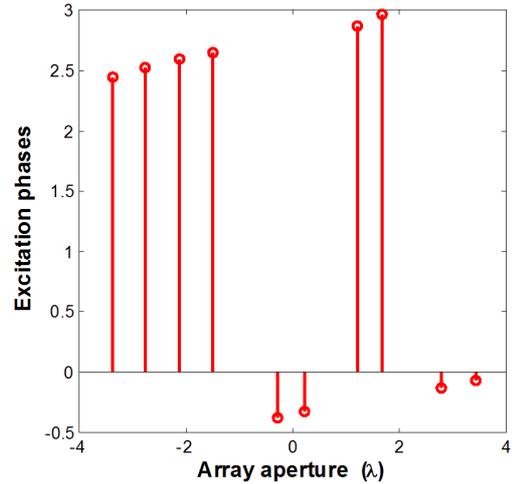

Fig. 11. Excitation phases corresponding to the red pattern of Fig. 8.

*IV.4 Comparison with [7] (cosecant beam)*

We adopted as reference power pattern the square-cosecant field shown in blue color in Fig. 12. In [7], by adopting BCS this pattern has been synthesized through a sparse array composed by 13 elements located over an aperture of 7.5λ. By exploiting the proposed approach, we have been able to reduce again the array's number of elements and size, respectively, without any radiation-performance loss.

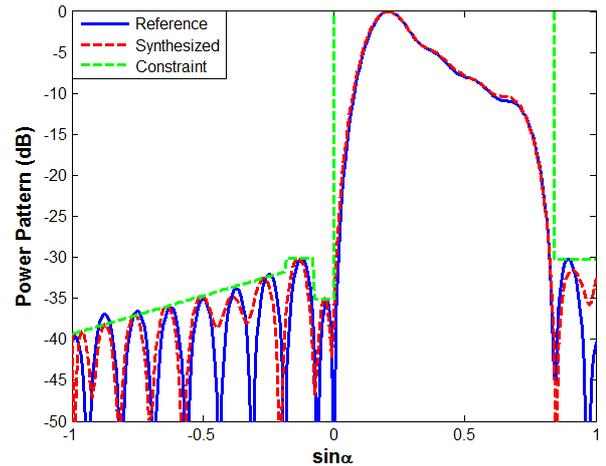

Fig. 12. Synthesis of a maximally-sparse array radiating a square-cosecant pattern with a specific sidelobes decay. Reference (blue curve, from [7]) and synthesized (red curve) square-amplitude array factors. Adopted upper bound (green curve) also depicted. Simulation parameters: $d/\lambda=0.5$; $Q=16$; $N=161$; $\nu=0.03$; $\sigma/\lambda=0.05$; $\eta/\beta=0.01$.

This circumstance is shown in Fig. 12, wherein a comparison between reference and synthesized patterns is reported. The designed array is composed by 12 elements located over an aperture of 6.9λ with a minimum spacing of 0.51λ. The corresponding excitation amplitudes and phases are depicted in figures 13 and 14, respectively.

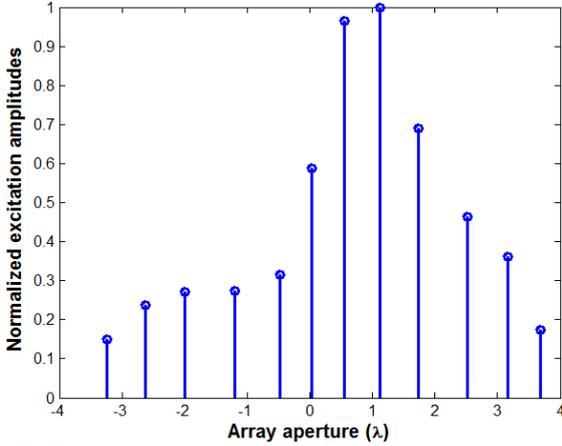

Fig. 13. Excitation amplitudes of the maximally-sparse array radiating the red pattern of Fig. 12.

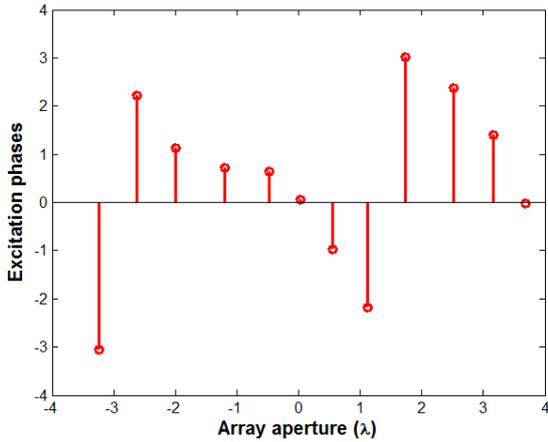

Fig. 14. Excitation phases of the maximally-sparse array radiating the red pattern of Fig. 12.

*IV.5 Comparison with [10] (cosecant beam)*
In the fifth test case, we aimed at synthesizing the square-cosecant power pattern depicted in blue color in Fig. 15. In [10], CS has been exploited to generate it by means of a sparse array composed by 13 elements located over an aperture of 8λ. Notably, by means of the presented approach, we have achieved an equivalent radiation performance while reducing both the elements number and the aperture size. In particular, the synthesized sparse array is composed by 12 elements located over an aperture of 7.5λ with a minimum spacing of 0.56λ.

The achieved antenna layout and excitations are depicted in Fig. 16 (amplitude distribution) and Fig. 17 (phase distribution), respectively. A superposition of the reference and synthesized patterns is shown in Fig. 15, wherein it can be noticed that all the design goals and constraints have been fulfilled. The same figure shows the upper-bound function exploited to bind the sidelobes as desired, which also points out the flexibility of the overall synthesis procedure.

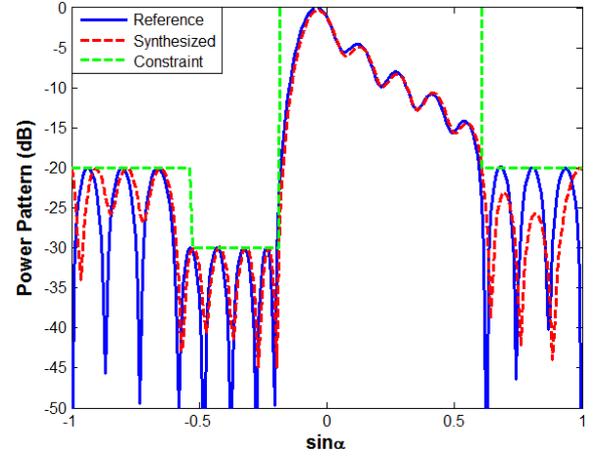

Fig. 15. Synthesis of a maximally-sparse array radiating a square-cosecant pattern with a non-uniform, piecewise-constant sidelobe level behavior. Adopted upper bound (green curve); reference (blue curve, from [10]) and synthesized (red curve) square-amplitude array factors. Simulation parameters: $d/\lambda=0.5$; $Q=16$; $N=161$; $\nu=0.005$; $\sigma/\lambda=0.05$; $\eta/\beta=0.08$.

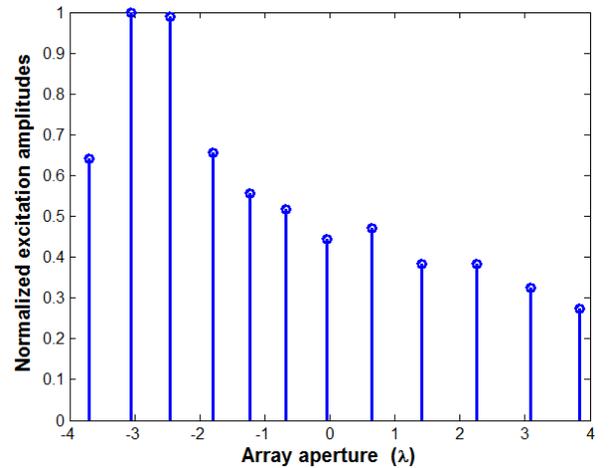

Fig. 16. Array excitation amplitudes corresponding to red pattern of Fig. 15.

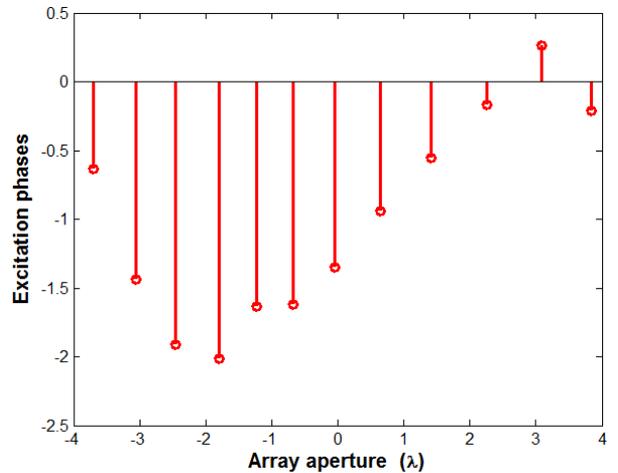

Fig. 17. Array excitation phases corresponding to the red pattern of Fig. 15.

*IV.6 Synthesis of a factorable beam through a planar array*

As a final test case, we show the outcomes of a numerical experiment concerning a simple way to extend the proposed procedure to the synthesis of planar sparse arrays with factorable patterns.

Coming to details, we aimed at synthesizing a beam having a flat-top behavior along the azimuth angle and a square-cosecant behavior along the elevation angle, which is of interest in radar applications as well as for radio-base stations. In so doing, we took as a reference the pattern generated from the factorization of the fields depicted in blue color in figures 8 and 12, respectively. Notably, exploitation of the results in [7] would require 13x11=143 elements to generate the desired radiation pattern. On the other side, a straightforward application of our results in Subsections IV.3 and IV.4 already reduces to 12x10=120 the number of required elements. This circumstance is coherent with the fact that, as long as a pattern is factorable, the elements number reduction in the planar array is roughly doubled with respect to the one experienced in the two underlying one-dimensional arrays.

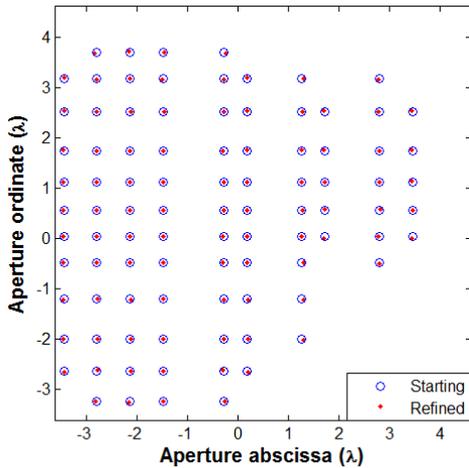

Fig. 18. Array layout achieved by factorizing the one-dimensional solutions shown in Subsections IV.3 and IV.4 and discarding those elements having a normalized excitation amplitude lower than 0.04: elements' location before (blue circles) and after (red dots) running the final 'refinement' algorithm.

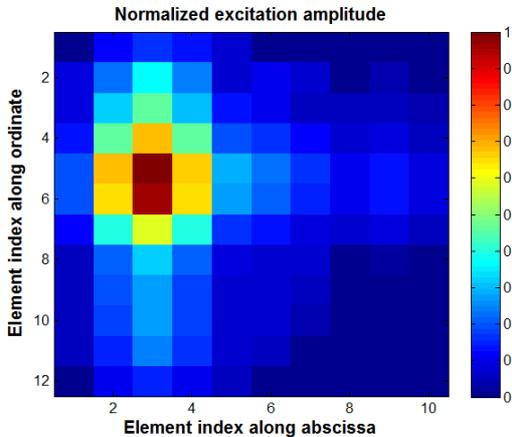

Fig. 19. Normalized excitation amplitudes associated to the refined layout shown in Fig. 18.

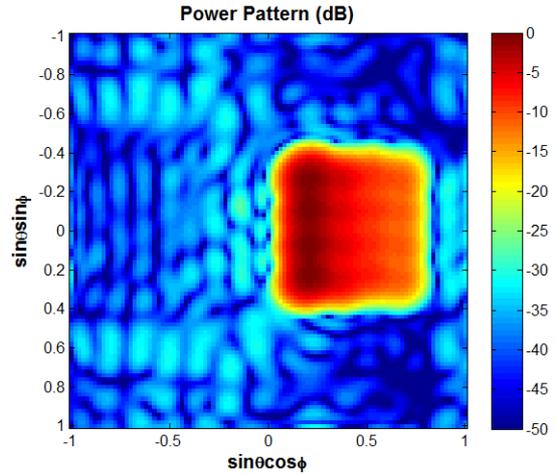

Fig. 20. Power pattern of the planar array having the refined layout of Fig. 18 and the excitations whose amplitude distribution is shown in Fig. 19.

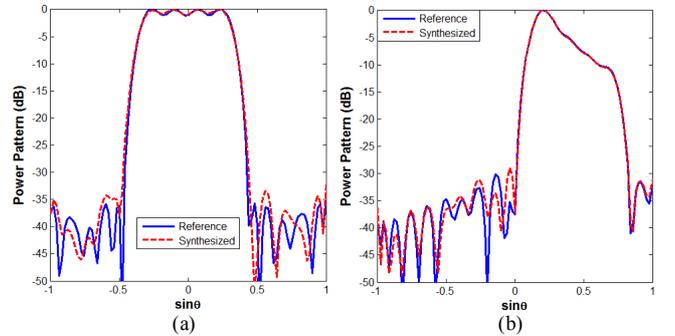

Fig. 21. Main cuts along ordinate [subplot (a)] and abscissa [subplot (b)] of the power pattern shown in Fig. 20: reference (blue color) and synthesized (red color) solutions.

Interestingly, a simple application of the third step of the proposed procedure, i.e., of the techniques presented in Subsection III.3, allows a further considerable reduction of elements. In fact, by simply 'eliminating' those elements having an excitation amplitude lower than $\upsilon=0.04$ we achieved the array layout depicted in Fig. 18 (blue circles), which is composed by 94 elements. Then, by adapting to the two-dimensional layouts case the approach in (10),(11), in such a way to *jointly* refine *both* the $x$ and $y$ array locations, we finally achieved the layout depicted in Fig. 18 (red dots), and the corresponding optimal excitations depicted in Fig. 19 (just amplitude is shown). The achieved power pattern is shown in figures 20 (in the two-dimensional spectral plane) and 21 (along its two main cuts), wherein $\theta$ and $\phi$ respectively represent the elevation and azimuth angles. As it can be seen, a very good agreement is achieved between the reference field and the main cuts of the synthesized pattern.

Notably, the overall approach allows to save roughly the 58% of the elements with respect to a fully populated array with a uniform $\lambda/2$ spacing, and roughly the 34% of elements (tolerating a slight worsening of the very low sidelobe levels) with respect to a simple factorization of the solutions in [7].

IV. CONCLUSIONS

We have introduced and tested a new approach, inspired from CS, for the synthesis of shaped beams by means of linear arrays having the minimum possible number of elements. The presented technique takes maximum advantage from the multiplicity of different far field and source distributions (having different amounts of sparsity) corresponding to a fixed power pattern, and also from the circumstance that different power patterns may equally fulfill the initial constraints.

These features, as well as the use of a reduced cardinality of the tentative array, has allowed us to improve the amount of sparsity of the synthesized array in a number of benchmark problems present in the state-of-the-art literature. In particular, different methods based on LP [14], FBMPM [12], CS [10] and BCS [7] have been considered for comparison. In all cases, the proposed approach always performed equally or better than the considered techniques.

The procedure can be easily extended to other cases of interest in several applications, including the design of arrays with factorable patterns (see Section IV.6) as well as one-dimensional simply reconfigurable arrays (by exploiting the ideas in [36]).